# Magnetic modulation doping in topological insulators toward higher-temperature quantum anomalous Hall effect

M. Mogi,[1,a)] R. Yoshimi,[1] A. Tsukazaki,[2,3] K. Yasuda,[1] Y. Kozuka,[1] K. S. Takahashi,[4] M. Kawasaki,[1,4] and Y. Tokura[1,4]

[1]*Department of Applied Physics and Quantum Phase Electronics Center (QPEC), University of Tokyo, Tokyo 113-8656, Japan*

[2]*Institute for Materials Research, Tohoku University, Sendai 980-8577, Japan*

[3]*PRESTO, Japan Science and Technology Agency (JST), Chiyoda-ku, Tokyo 102-0075, Japan*

[4]*RIKEN Center for Emergent Matter Science (CEMS), Wako 351-0198, Japan*

**Quantum anomalous Hall effect (QAHE), which generates dissipation-less edge current without external magnetic field, is observed in magnetic-ion doped topological insulators (TIs), such as Cr- and V-doped $(Bi,Sb)_2Te_3$. The QAHE emerges when the Fermi level is inside the magnetically induced gap around the original Dirac point of the TI surface state. Although the size of gap is reported to be about 50 meV, the observable temperature of QAHE has been limited below 300 mK. We attempt magnetic-Cr modulation doping into topological insulator $(Bi,Sb)_2Te_3$ films to increase the observable temperature of QAHE. By introducing the rich-Cr-doped thin (1 nm) layers at the vicinity of the both surfaces based on non-Cr-doped $(Bi,Sb)_2Te_3$ films, we have succeeded in observing the QAHE up to 2 K. The improvement in the observable temperature achieved by this modulation-doping appears to be originating from the suppression of the disorder in the surface state interacting with the rich magnetic moments. Such a superlattice designing of the stabilized QAHE may pave a way to dissipation-less electronics based on the higher-temperature and zero magnetic-field quantum conduction.**

---

a) Electronic mail: mogi@cmr.t.u-tokyo.ac.jp

The three-dimensional topological insulators (TIs) with the energy gap of the bulk state have spin-momentum locked massless Dirac electrons at the surface as protected by time reversal symmetry.[1,2] By breaking time reversal symmetry, e.g. via ferromagnetism, the Dirac energy dispersion acquires mass-gap. As the top and bottom surface states in thin films are connected by the insulating bulk and the localized state on the narrow side walls, each surface state contributes a half-quantized value of $e^2/2h$ consequently when Fermi energy ($E_F$) locates in the induced gap of the surface state.[3] Under this situation, the Hall conductance with the integer quantized values of 0 or ±1 times $e^2/h$ emerges even without external magnetic field,[4-13] namely quantum anomalous Hall (QAH) effect. The QAH effect is intriguing also for possible application of dissipation-less edge current as controlled by the magnetic domains.

Following the experimental discovery of QAH effect by Chang *et al.*,[6] the QAH effect has been reproducibly observed and reported for the molecular-beam-epitaxy (MBE) grown (Bi,Sb)$_2$Te$_3$ thin films doped with magnetic elements such as Cr[6-11] and V.[12,13] However the observable temperature of the QAH effect has been limited below 300 mK, typically at a dilution refrigerator, which is two orders of magnitude lower than the Curie temperature in spite of the large magnetically induced gap of about 50 meV.[14,15] The main reasons for this discrepancy have not been clarified, but we speculate that those are (1) inhomogeneity of coupling between the surface Dirac electron and the doped magnetic moments and (2) presence of the parasitic conduction at the bulk region formed by the inhomogeneity of Bi/Sb and Cr. The problem (1) leads to spatial fluctuation of magnetic mass-gap of the surface state as partly probed by scanning tunneling spectroscopy.[15] This fluctuation diminishes the effective gap magnitude, therefore the QAH state is adapted to be easily broken by thermal activation for the electrical observation. The problem (2) yields the residual longitudinal resistance originating from the mixing with dissipative bulk conductive channels when rich-Cr is doped into the (Bi,Sb)$_2$Te$_3$.[7] Therefore, rich-Cr doped single-layer does not naively become the best platform for higher-temperature QAH effect. These problems must be addressed to realize the robust QAH effect and the exploratory researches of exotic functions based on QAH effect.

In this Letter, we report the observation of QAH effect at 1 K (and possibly subsisting up to 2 K) by applying ferromagnetic (Cr) modulation-doping to topological insulator $(Bi_{1-y}Sb_y)_2Te_3$ (BST) thin films. In conventional modulation doping, a donor element is introduced apart from a 2D conducting channel at the heterointerface such as $GaAs/Al_{1-x}Ga_xAs$ to reduce ionized-impurity scattering, as contrasted by uniform doping.[16,17] In the present magnetic modulation-doping, high concentration of Cr doped monolayers are introduced to the vicinity of the surface to bestow the ferromagnetism on the spin-polarized surface state and to enhance the effective mass-gap with suppression of disorder.[18]

The films were grown in a MBE chamber on semi-insulating InP(111) substrates according to the same procedures described in Ref. 7 except for a growth temperature (200 ºC). The total thickness of films was fixed to 8 nm with the deposition rate (~ 0.2 nm/min) determined from X-ray diffraction measurement in uniformly doped samples. In synthesis, Bi, Sb and Te were co-evaporated and Cr was selectively supplied for Cr modulation-doping. It was confirmed in our previous study[18] that the Cr could be doped in selective layers without inter-diffusion. To tune the chemical potential at the charge neutral point, we varied the Bi/Sb composition ratio ($y$) controlled by beam equivalent pressure of Bi and Sb, and fabricated top-gated field effect transistor devices of Hall-bar geometry with 33-nm $AlO_x$ gate dielectric layer. Low-temperature transport measurements were carried out in a Quantum Design Physical Property Measurement System (PPMS) using $^3$He system above 0.5 K and in a dilution refrigerator below 0.5 K.

To explore the optimal structure of the modulated TI films for stabilization of QAH effect, we fabricated three devices as shown in Figs. 1(a)-(c). The Bi/Sb ratio ($y$) was fixed at 0.78. Figure 1(a) shows a uniformly doped $Cr_x(Bi_{1-y}Sb_y)_{2-x}Te_3$ (CBST) ($x = 0.10$) film ("single-layer") which shows QAH effect at 50 mK. Figures 1(b) and 1(c) show the schematic structures with magnetic modulation doping;

in the "tri-layer" structure (Fig. 1 (b)) the top and bottom 1-nm surface layers are intensively Cr-doped ($x$ = 0.46), while in the "penta-layer" structure (Fig. 1(c)) Cr ($x$ = 0.46) was doped in the similar way apart, but by 1 nm away from the topmost and bottommost surfaces. The thickness of respective layers was nominally determined from the deposition rate. The Curie temperature of single-, tri- and penta-layer are about 9 K, 25 K and 25 K, respectively estimated from the temperature dependence of $R_{yx}$ under zero magnetic fields (see the Fig. 2(d) and the inset of it). Note that the nominal integrated Cr amounts in three films were tuned to almost the same values and that Cr-doping affects the chemical potential in CBST.

We show in Figs. 1(d) and 1(e) the gate voltage ($V_G$) dependence of Hall resistance ($R_{yx}$) and longitudinal resistance ($R_{xx}$) at 0.5 K without external magnetic field ($B$ = 0 T) after magnetic-field cooling ($B$ = 2 T). The charge neutral points $V_{CNP}$, where Hall angle ($\tan^{-1}(R_{yx}/R_{xx})$) gets a maximum value, are 1.5 V for single-layer, 0.9 V for tri-layer and 5.5 V for penta-layer. In the single-layer (gray line), $R_{yx}$ is about 0.85 $h/e^2$ and $R_{xx}$ has a dip at $V_G = V_{CNP}$, indicating the incipient feature of the metallic chiral edge state in QAH regime. The values of $R_{yx}$ and $R_{xx}$ at 0.5 K is comparable to the previous single-layer studies.[6,7] The tri-layer (blue line) has the similar gate voltage dependence and approaches more closely to the QAH state than the single-layer case. In the penta-layer (red line), $R_{yx}$ shows the quantized Hall plateau ($h/e^2$) at around $V_G = V_{CNP}$, indicating the occurrence of QAH effect even at 0.5 K. Figures 1(f) and 1(g) show the $V_G$ dependence at the lowest temperature of the present study, 50 mK. At $V_G = V_{CNP}$, all three devices show the well-defined QAH state, i.e. $R_{yx} = h/e^2$ and $R_{xx} \sim 0$ Ω. Nevertheless, a remarkable difference is seen in the width of the QAH plateau against $V_G$ variation; the wide plateau in $V_G$ variation probably reflects (1) the large effective mass-gap in modulation-doped structure and (2) difficulty of $E_F$ tuning via gating. These two possibilities originate from the rich-Cr-doped layer providing larger mass-gap and parasitic conduction in 1-nm layer, but the adverse implication is effectively minimized by the modulated structure.

To clarify the characteristics of QAH effect in each structure, we investigate the magnetic field $B$ and temperature $T$ dependence under $V_G = V_{CNP}$. Figures 2(a), 2(b) and 2(c) show the $B$ dependence of $R_{yx}$ and $R_{xx}$ at 0.5 K in the single-layer, tri-layer and penta-layer structures, respectively. Ferromagnetic hystereses with the out-of-plane easy axis are seen in $R_{yx}$ (blue line) for all the structures. At the coercive field ($\mu_0 H_c$) corresponding to the plateau-to-plateau transition, $R_{xx}$ (red line) exhibits a peak, which probably implies that the domain walls within the small multi-domains provide an objective effect to the edge current conduction.[19] The single-layer configuration (Fig. 2(a)) shows appreciable deviations from the quantized value for $R_{yx}$ or zero for $R_{xx}$ at 0.5 K, while showing the well-defined QAH state at 50 mK (see the inset of Fig. 2(a)). Contrary to the single-layer, $R_{yx}$ for the tri-layer and for the penta-layer almost reaches the quantized values of $\pm h/e^2$ at 0.5 K. Note that the $V_G$ dependence of $R_{yx}$ at $B = 0$ T shown in Fig. 1(d) did not reach the quantized value for the tri-layer structure because of the coexistence of the anti-domain state in the zero magnetic field. In particular, the penta-layer structure shows the nearly ideal QAH effect where $R_{yx} = \pm 1.00\ h/e^2$ and $R_{xx} = 0.058\ h/e^2$, even at 0.5 K. Figures 2(d) and 2(e) show the $T$ dependence of $R_{yx}$ and $R_{xx}$ in the single-layer (gray), tri-layer (blue) and penta-layer (red), respectively; there, a low magnetic field (0.2 T) was applied in order to ensure the formation of a single magnetic domain state. As temperature decreases, $R_{yx}$ increases below Curie temperature because the magnetization becomes large and the magnetic mass-gap opens. In the order of single-, tri-, and penta-layer, the temperature at which $R_{yx}$ approaches the quantized value is getting higher. Another litmus test for the QAH state is the localization of surface-state electrons in QAH regime, which can be characterized by the $T$ dependence of $R_{xx}$. As lowering temperature, $R_{xx}$ first increases and then decreases toward zero. The most notable trend is that the starting temperature of localization (triangles in Fig. 2(e)) is getting higher in the order of single-, tri-, and penta-layer; the temperature for single-layer is ~ 1 K (not 6 K because the temperature is close to Curie temperature), ~ 7 K for tri-layer, and ~ 15 K for penta-layer. From these results, it is concluded that the penta-layer is a most suitable structure for observing higher-temperature QAH effect.

Toward higher temperature realization of QAH effect based on penta-layer, we have exemplified three penta-layer devices with varying the Cr concentration ($x$). Accompanying with the $x$ variation, Bi/Sb ratio ($y$) was also slightly tuned so that $V_{CNP}$ can be detected in $V_G$ sweep range. The tested compositions in the penta-layer are ($x$, $y$) = (0.46, 0.78), (0.57, 0.74), (0.95, 0.68). The results for the ($x$, $y$) = (0.46, 0.78) were already shown in Figs. 1 and 2. The respective Curie temperatures are about 25 K, 55 K and 80 K that is judged from $T$ dependence of $R_{yx}$. Figures 3(a) and 3(b) show the $V_G$ dependence of $R_{yx}$ and $R_{xx}$ at $T$ = 0.5 K and $B$ = 0 T. Among them, the ($x$, $y$) = (0.57, 0.74) penta-layer presents a wide quantized Hall plateau in Fig. 3(a) and a minimum value of $R_{xx}$ in Fig. 3(b). The highest Cr-doped ($x$, $y$) = (0.95, 0.68) penta-layer shows worse behavior of $R_{yx}$ and $R_{xx}$ in whole $V_G$ region far from QAH state. Figures 3(c) and 3(d) show the $B$ dependence of them at 0.5 K under $V_G$ = $V_{CNP}$. It is to be noted that the higher Cr concentration provides larger coercive field ($\mu_0 H_c$) with broad hysteresis. Thus there is an optimum Cr concentration in the penta-layer in this present study, ($x$, $y$) = (0.57, 0.74), presenting a well-defined QAH effect at 0.5 K, where the residual $R_{xx}$ at zero magnetic field is as small as 0.017 $h/e^2$.

In Figs. 4, the magnetic field dependence of $R_{yx}$ and $R_{xx}$ for the presently optimized penta-layer ($x$, $y$) = (0.57, 0.74) under $V_G$ = $V_{CNP}$ is compared with varying temperature; 0.5 K, 1 K, 2 K, and 4.2 K. Quantized Hall resistance is observed up to 1 K, where the residual longitudinal resistance is 0.081 $h/e^2$, which is a little higher than that at 0.5 K (0.017 $h/e^2$). At 2 K, $R_{yx}$ is still close to the quantized value, ±0.97 $h/e^2$. Then at 4.2 K, it goes away from QAH state ($R_{yx}$ ~ ±0.87 $h/e^2$), yet showing the large Hall angle, 66.1°. To observe QAH effect beyond liquid $^4$He temperature, further research is required, for example, fine tuning of Cr concentration ($x$) and Bi/Sb ratio ($y$) to suppress disorder and/or doping another magnetic element such as V which can raise the Curie temperature as compared with the similar concentration of Cr.[12]

As a final remark, we discuss the origin for the increase in observable temperature of QAH effect in the magnetic modulation doped structures. Since the Hall conductivity in QAH regime is successfully observed with a definite quantized value of ± $e^2/h$, i.e. twice a half quantized value, it is likely that the modulation-doping at the vicinity of the top and bottom surfaces keeps the nature of the three-dimensional topological insulator for the whole film. In the ideal QAH state, there is no bulk conducting state in the surface mass-gap $\Delta \propto JM_z$, where $J$ is the exchange interaction between surface electrons and magnetic dopant atoms and $M_z$ is the magnetization perpendicular to the surface.[5,20] In real electrical measurements, however, the size of mass-gap is effectively reduced by inhomogeneity of Cr concentration and bulk conducting states. Since the $M_z$ is proportional to Cr concentration, the effective mass-gap should be subject to spatial fluctuation in magnetic doped layer, which diminishes the observable temperature of QAH effect far below the Curie temperature. Magnetic modulation doping makes it possible to dope high concentration of Cr and provide strong magnetic coupling between the surface state and the near-by magnetic moments suppressing bulk conduction, which leads to acquisition of large effective mass-gap. This is why the observable temperature is higher in tri-layer and penta-layer than single layer. In contrast, the strength of exchange coupling ($J$) in the penta-layer would be smaller than in the tri-layer, considering the doped position of magnetic moments.[21] In Cr-doped $(Bi,Sb)_2Te_3$ films, the ferromagnetic interaction itself is primarily mediated not by the Dirac electrons (RKKY interaction) but by the bulk property, i.e., the van Vleck mechanism.[4,22,23] Therefore, the difference in the mass-gap would come from the difference in the $J$ not $M_z$. However, the QAH state in the penta-layer structure is more stabilized than that in the tri-layer, as shown in Figs. 1 and 2, while their Curie temperatures are comparable. Therefore, we can conclude that the important factor to enhance the thermal stability of the QAH state is not only the size of effective mass-gap but also its spatial homogeneity, which is improved by the remote doping of the magnetic moments from the surface layer in penta-layer structure.

In conclusion, we have investigated the effects of magnetic modulation doping into topological insulators. By inserting the rich-Cr-doped 1-nm-thick layers locating at 1 nm beneath and above the top and bottom surfaces in the chemical-potential tuned $(Bi,Sb)_2Te_3$ (BST) films with the total thickness of 8 nm, we successfully stabilized and observed the clear QAH effect up to 1 K, more than an order of magnitude higher than that for the uniformly Cr-doped BST single-layer film. The impact of the modulation doping is the enhancement of magnetically induced mass-gap because of higher Cr doping concentration as well as the reduction of the doping induced disorder in the surface-state conduction. The present method based on the topological insulator superlattice engineering[18,24,25] would pave a way to deepened researches on zero magnetic-field dissipation-less edge current and further to exploration of exotic topological phenomena.

The authors thank N. Nagaosa and M. Kawamura for enlightening discussions. This research was supported by the Japan Society for the Promotion of Science through the Funding Program for World-Leading Innovative R&D on Science and Technology (FIRST Program) on "Quantum Science on Strong Correlation" initiated by the Council for Science and Technology Policy and by JSPS Grant-in-Aid for Scientific Research (S) No.24224009 and 24226002.

**FIGURES**

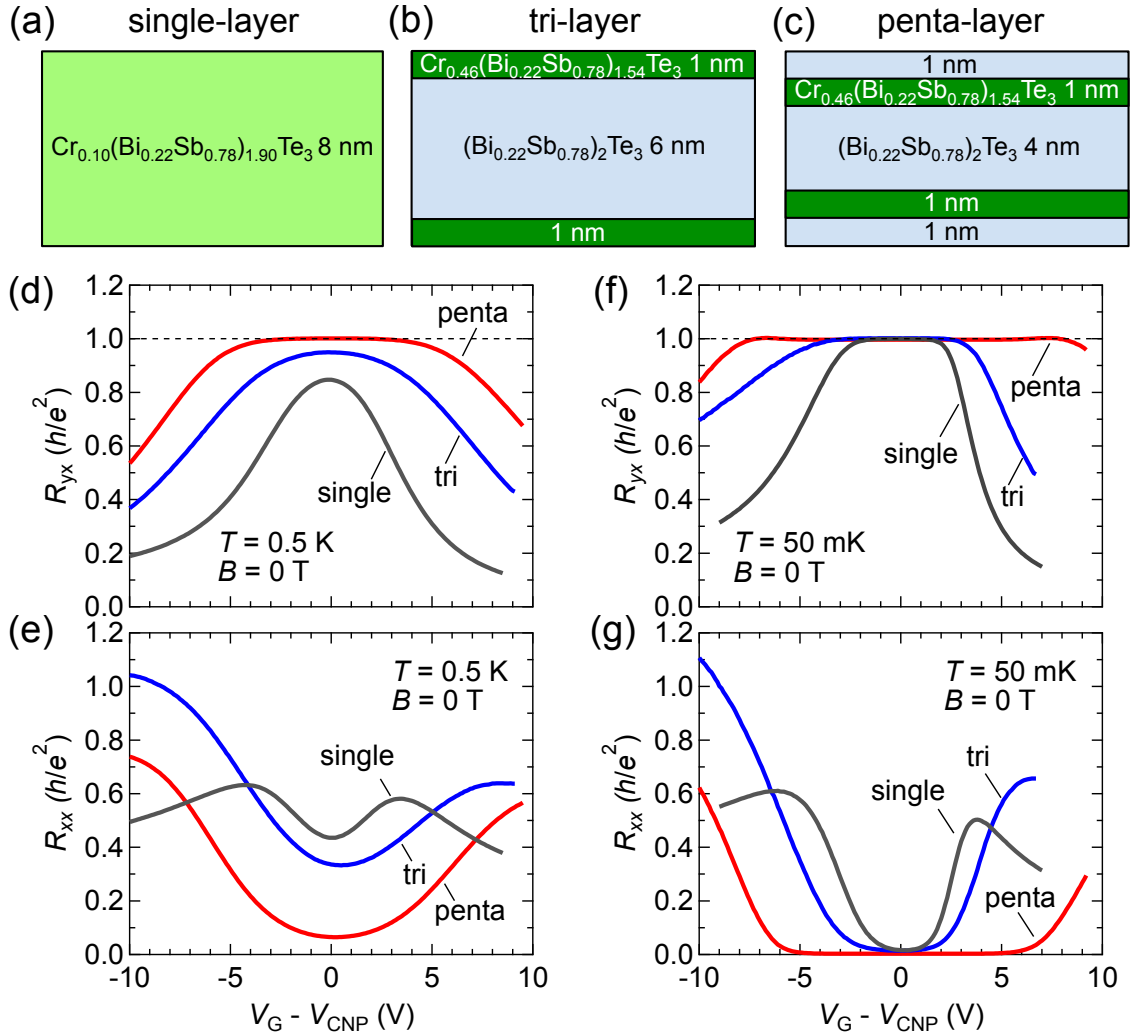

FIG. 1. Schematics of the tested uniform and modulation magnetic doping structures; (a) $Cr_x(Bi_{1-y}Sb_y)_{2-x}Te_3$ ($x = 0.10$, $y = 0.78$) single-layer. (b) tri-layer ($x = 0.46$, $y = 0.78$), and (c) penta-layer ($x = 0.46$, $y = 0.78$). Gate voltage ($V_G$) dependence of (d,f) Hall resistance ($R_{yx}$) and (e,g) longitudinal resistance ($R_{xx}$) of the single-layer (gray line), the tri-layer (blue line) and the penta-layer (red line) at 0.5 K and at 50 mK, respectively, with no external magnetic field ($B = 0$ T).

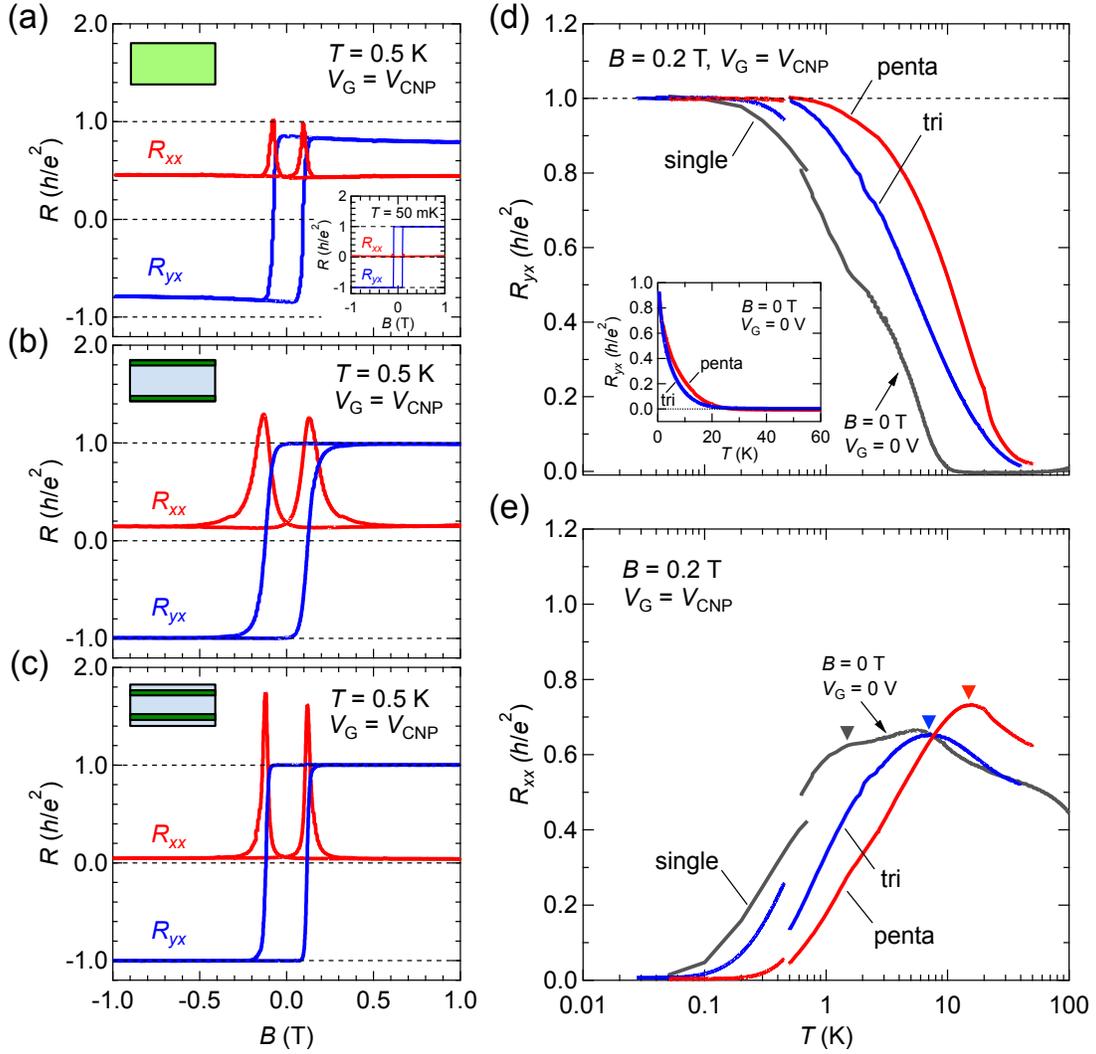

FIG. 2. Magnetic field ($B$) dependence of Hall resistance ($R_{yx}$: blue line) and longitudinal resistance ($R_{xx}$: red line) in (a) the single-layer, (b) the tri-layer, (c) the penta-layer at 0.5 K, tuned at the charge neutral point by gating. The inset to (a) shows the result of the single-layer at 50 mK. Temperature ($T$) dependence of (d) $R_{yx}$ and (e) $R_{xx}$ in the respective structures with magnetic field $B = 0.2$ T. Triangles in (e) show the starting points of localization in the respective structures. In the single-layer, gate voltage ($V_G$) and magnetic field are not applied. The inset to (d) shows the behaviors for the tri-layer and penta-layer films without applying gate voltage and magnetic field (plotted on a linear scale).

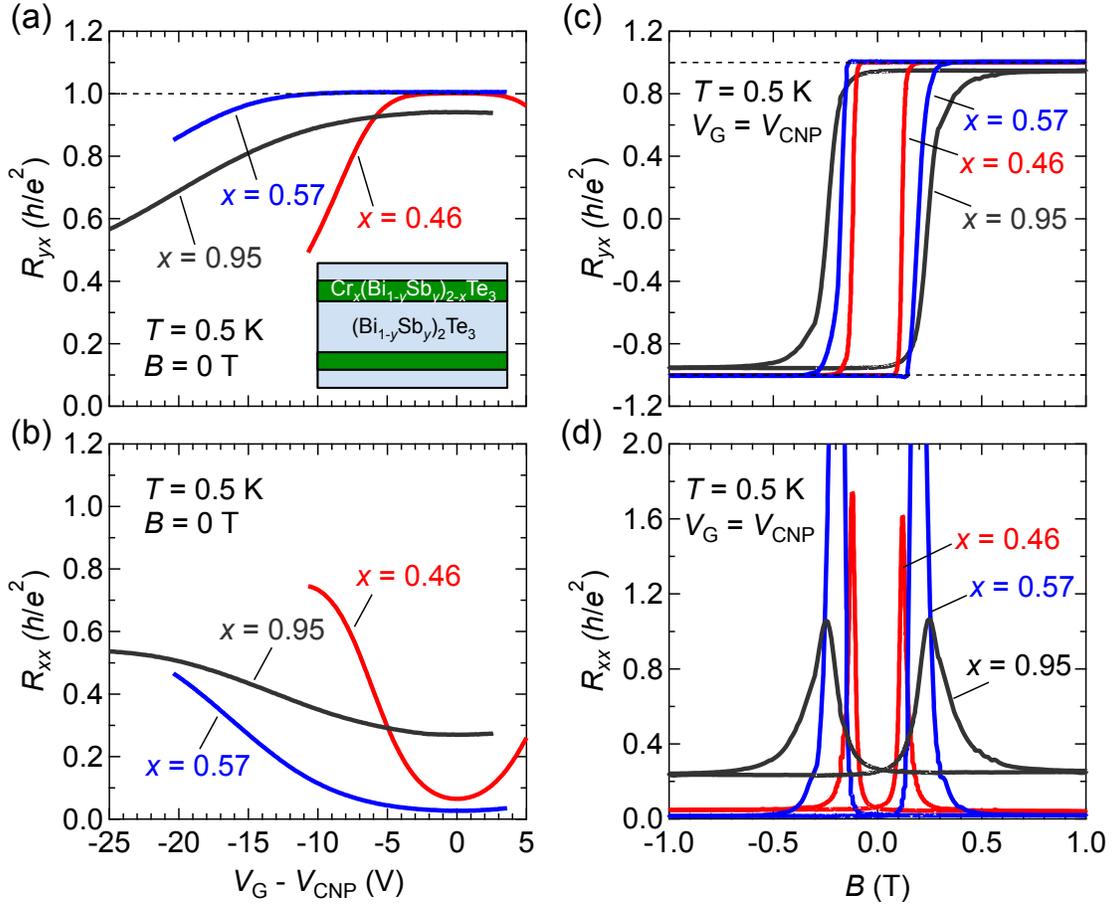

FIG. 3. Gate voltage dependence ($V_G$) of (a) Hall resistance ($R_{yx}$) and (b) longitudinal resistance ($R_{xx}$) in the penta-layer devices at 0.5 K without magnetic field ($B$ = 0 T). The results for three devices with the different compositions in the penta-layer structure are shown; ($x$, $y$) = (0.46, 0.78) (red line), (0.57, 0.74) (blue line), and (0.95, 0.68) (gray line). Magnetic field ($B$) dependence of (c) $R_{yx}$ and (d) $R_{xx}$ in the respective compositions at 0.5 K, tuned at the charge neutral point by gating ($V_G = V_{CNP}$).

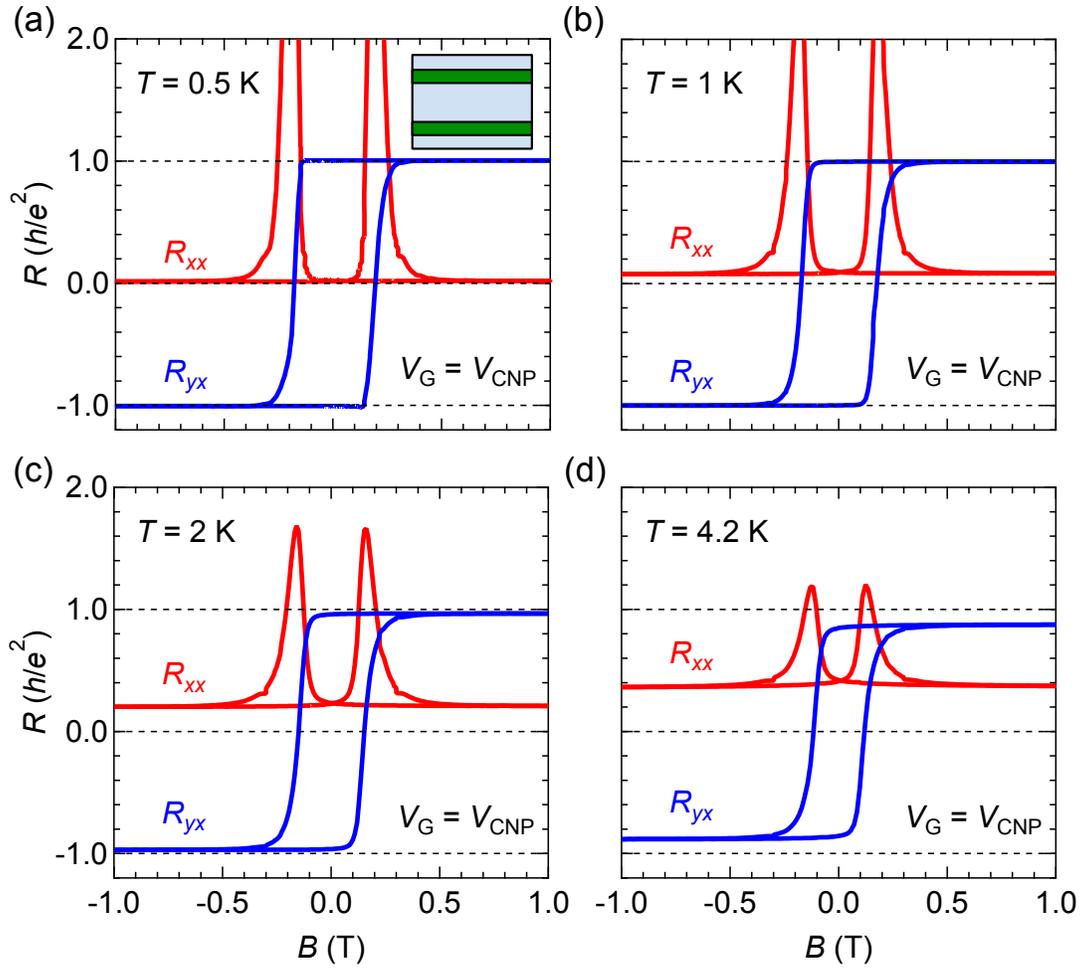

FIG. 4. Magnetic field ($B$) dependence of Hall resistance ($R_{yx}$) and longitudinal resistance ($R_{xx}$) in the penta-layer device ($x = 0.57$, $y = 0.74$) at 0.5 K (a), 1 K (b), 2 K (c) and 4.2 K (d). Gate voltage ($V_G$) was tuned at the charge neutral point ($V_{CNP}$).